\documentclass[review,3p]{elsarticle}
\biboptions{sort&compress}

\usepackage{color}
\usepackage{amssymb, amsmath}
\usepackage{multirow}

\usepackage{algorithm}
\usepackage[noend]{algpseudocode}
\algblockdefx{ParallelForEach}{EndParallelForEach}[1]{\algorithmicfor~\textbf{each} #1 \textbf{in parallel}}{\textbf{end parallel for each}}
\makeatletter
\ifthenelse{\equal{\ALG@noend}{t}}{\algtext*{EndParallelForEach}}{}
\makeatother

\newcommand{\vv}[1]{\mathbf{#1}}

\journal{Computer Physics Communications}

\begin{document}
\begin{frontmatter}
\title{Efficient mesoscale hydrodynamics: multiparticle collision dynamics with massively parallel GPU acceleration}
\author[princeton]{Michael P. Howard}
\author[princeton]{Athanassios Z. Panagiotopoulos}
\author[mainz]{Arash Nikoubashman\corref{corrauth}}
\address[princeton]{Department of Chemical and Biological Engineering, Princeton University, Princeton,
                    New Jersey 08544, United States}
\address[mainz]{Institute of Physics, Johannes Gutenberg University Mainz, Staudingerweg 7,
                55128 Mainz, Germany}
\cortext[corrauth]{Corresponding author}


\begin{abstract}
We present an efficient open-source implementation of the multiparticle
collision dynamics (MPCD) algorithm that scales to run on hundreds of graphics
processing units (GPUs). We especially focus on optimizations for modern GPU
architectures and communication patterns between multiple GPUs. We show that a
mixed-precision computing model can improve performance compared to a
fully double-precision model while still providing good numerical accuracy.
We report weak and strong scaling benchmarks of a reference MPCD solvent and
a benchmark of a polymer solution with research-relevant interactions and system
size. Our MPCD software enables simulations of mesoscale hydrodynamics at length
and time scales that would be otherwise challenging or impossible to access.
\end{abstract}
\end{frontmatter}

\section{Introduction}
Complex fluids and soft matter, including colloidal suspensions, polymer
solutions, and biological materials, are characterized by
structure on disparate length scales, dynamics that occur over a range of time
scales, and interactions on the scale of thermal energy \cite{Praprotnik:2008er,Nagel:2017jm}. This combination of
properties can give rise to many complex behaviors, including non-Newtonian
rheology \cite{Brady:1993vs,Phung:1996wh,Winkler:2013kx} and the ability to
self-assemble into organized structures \cite{Whitesides:2002uq,Grzybowski:2009ke}. Computer
simulations have emerged as useful tools for studying complex fluids and soft
matter because they offer detailed, simultaneous resolution of structure
and dynamics, and can be used to predict and systematically engineer
a material's properties.

Many complex fluids and soft materials consist of a mesoscopic solute, such
as polymer chains or spherical colloids, dispersed in a molecular solvent. The solute
can be orders of magnitude larger than the solvent and have
corresponding slower dynamics \cite{Padding:2006uz}. For example, colloidal particles typically
have diameters ranging from a few nanometers to micrometers, but are dispersed in
solvents such as water that have molecular diameters less than a nanometer.
Classical molecular dynamics (MD) simulations \cite{Allen:1991,Frenkel:2002chF} are often unsuitable for such
problems because explicitly resolving the solvent with the solute quickly becomes
intractable, both in terms of the number of particles required for the model
and the number of simulation time steps that are necessary to study the
time scales of interest. On the other hand, continuum constitutive models
employed in computational fluid dynamics \cite{Boger:1977vy,Giesekus:1982vs} neglect details of
fluctuating microscopic structures, which may be important for accurately describing
complex liquids such as polymer solutions when the rheological properties are influenced
by the local structure. A simulation
approach bridging these regimes to retain a detailed model of the solute
while resolving only the essential effects of the solvent is necessary.
Various mesoscale methods have been developed for this purpose, including
lattice-based models such as the family of Lattice-Boltzmann methods \cite{Chen:1998vva,Ladd:2001gv,Duenweg:2009ie}; implicit-solvent
models including Brownian dynamics \cite{Allen:1991}, Stokesian dynamics \cite{BRADY:1988up}, and fast lubrication
dynamics \cite{Kumar:2010hy}; and particle-based methods such as dissipative particle dynamics \cite{Hoogerbrugge:1992hl,Groot:1997du,Espanol:2017wy},
direct simulation Monte Carlo \cite{Bird:1963ef,Bird:1970bf},
and multiparticle collision dynamics (MPCD) \cite{Malevanets:1999wa,Gompper:2009is}.
The reader is referred to \cite{Bolintineanu:2014dj} and the references therein for
an excellent discussion on the various advantages and disadvantages of these methods,
particularly as applied to colloidal suspensions. This article focuses on the MPCD
method, first developed by Malevanets and Kapral \cite{Malevanets:1999wa}, and its
massively parallel implementation for graphics processing units (GPUs).

In MPCD, the solvent is represented as a set of off-lattice point particles that
undergo ballistic streaming steps followed by cell-based, stochastic multiparticle
collisions \cite{Malevanets:1999wa,Kapral:2008te,Gompper:2009is}. The nature and frequency of the
collisions determines the transport coefficients of the solvent \cite{Allahyarov:2002hq,Ihle:2003cn,Ripoll:2005ev}. The MPCD algorithm is unconditionally numerically
stable, has an H-theorem, and naturally includes the effects of thermal
fluctuations \cite{Malevanets:1999wa}. Polymers \cite{Malevanets:2000tm,Mussawisade:2005tj},
colloids \cite{Hecht:2005kz,Padding:2005ez,Poblete:2014gc,Bolintineanu:2014dj}, and fluid-solid boundaries \cite{Lamura:2001un,Whitmer:2010kp,Bolintineanu:2012ik} can be coupled to
the solvent, making MPCD an excellent tool for studying soft matter \cite{Gompper:2009is,Kanehl:2017fz,McWhirter:2009ht,Nikoubashman:2014en,Nikoubashman:2014vh,Howard:2015bl,Bianchi:2015dx,Bianchi:2015corr,Nikoubashman:2017bl}. The MPCD
equations of motion are less computationally demanding to integrate than those
in MD because MPCD particles do not usually have interactions with each
other (e.g., dispersion forces), which typically require some of the most
time-consuming calculations in the MD algorithm \cite{Allen:1991,Frenkel:2002chF}.
On the other hand, MPCD can require a high particle density in order to achieve a
sufficiently liquid-like solvent \cite{Ripoll:2005ev}, which still poses significant demands on the
memory required to represent the particle configuration and the number of
calculations required to propagate the solvent particles.

MPCD is particularly well-suited for parallelization
because of its particle-based nature and spatially localized collisions.
Yet, to our knowledge, there are few publicly available massively parallel implementations
of the MPCD algorithm. RMPCDMD \cite{deBuyl:2017hl} is a multithreaded program for performing
MPCD simulations, but lacks support for decomposing the simulation onto
multiple CPUs, limiting the simulation sizes that can be reasonably accessed.
The SRD package \cite{Petersen:2010cg} of the massively parallel MD code LAMMPS \cite{Plimpton:1995fc} implements a limited set
of MPCD features and supports spatial decomposition onto multiple CPU processes
using the Message-Passing Interface (MPI). Recently, Westphal et al. \cite{Westphal:2014db} described
an implementation of the MPCD algorithm for GPUs.
GPUs have become indispensable for performing MD simulations at large scale
because of their massively parallel architectures, and essentially all major MD
software packages support GPU acceleration to some extent. Westphal et al.
demonstrated a 1-2 order of magnitude acceleration of their MPCD code on a
GPU compared to single-threaded CPU code \cite{Westphal:2014db}. However, their implementation,
which to our knowledge is not publicly available, was designed to use a single
GPU as an accelerator, and so the GPU memory capacity effectively limits the size
of the simulations that can be performed. It is critical to be able to utilize
multiple GPUs through MPI in order to efficiently perform large-scale MPCD simulations.

In this article, we present a high-performance, open-source implementation of
the MPCD algorithm within the HOOMD-blue simulation package \cite{Anderson:2008vg,Glaser:2015cu} that scales to run
on hundreds of NVIDIA GPUs. We first give an overview of the MPCD algorithm
(Section~\ref{sec:alg}). We then describe the implementation of this algorithm with a focus
on optimization for the GPU and for supporting multiple GPUs with MPI
(Section~\ref{sec:implement}). We assess the accuracy and performance of a mixed-precision
computing model for the MPCD solvent. We report weak and strong scaling
benchmarks of a reference MPCD solvent and a benchmark of a polymer solution
(Section~\ref{sec:perf}).  We demonstrate that massively parallel GPU acceleration can enable
simulations of mesoscale hydrodynamics at length and time scales that would be
otherwise challenging or impossible to access.

\section{Algorithm\label{sec:alg}}
In MPCD \cite{Malevanets:1999wa,Kapral:2008te,Gompper:2009is}, the solvent is modeled by point particles of mass $m$ with continuous
positions $\vv{r}_i$ and velocities $\vv{v}_i$. MPCD particles are propagated through
alternating streaming and collision steps. During the streaming step, the
particles follow Newton's equations of motion. In the absence of any external
forces, the positions are simply updated during an interval of time $\Delta t$
according to:
\begin{equation}
\vv{r}_i(t+\Delta t) = \vv{r}_i(t) + \vv{v}_i(t) \Delta t.\label{eq:stream}
\end{equation}
The time step $\Delta t$ effectively sets the mean free path for the particles,
$\lambda = \Delta t \sqrt{k_{\rm B}T/m}$, where $k_{\rm B}$ is Boltzmann's
constant and $T$ is the temperature. Particles are then binned into cubic cells
of edge length $a$, which sets the length scale over which hydrodynamics
are resolved \cite{Huang:2012ev}. For values of $\lambda$ much smaller than $a$, the MPCD algorithm
violates Galilean invariance, which can be restored by applying a random shift
to the particle positions of $\pm a/2$ before binning \cite{Ihle:2001ty,Ihle:2003bq}. All particles in a cell
undergo a stochastic collision that updates their velocities while conserving
linear momentum. (Collision rules can be extended to also enforce angular-momentum
conservation \cite{Gotze:2007ec,Gompper:2009is}.) These stochastic collisions lead to a build up of hydrodynamic
interactions, and the choice of collision rule and solvent properties determine
the transport coefficients of the fluid.

In the stochastic rotation dynamics (SRD) collision rule \cite{Malevanets:1999wa},
the center-of-mass velocity of each cell, $\vv{u}$, is first computed and the relative
particle velocities are rotated by a fixed angle $\alpha$ about a randomly
chosen axis for each cell,
\begin{equation}
\vv{v}_i(t+\Delta t) = \vv{u}(t) + \vv{R}(\alpha) \cdot (\vv{v}_i(t) - \vv{u}(t)),\label{eq:srd}
\end{equation}
where $\vv{R}(\alpha)$ is the rotation matrix for the cell. Malevanets and Kapral
showed that the SRD rule yields the hydrodynamic equations for compressible flow,
has an H-theorem, and gives the correct velocity distribution for the particles \cite{Malevanets:1999wa}.
Although SRD in this form conserves linear momentum, eq.~\ref{eq:srd}
requires modification in order to also enforce angular-momentum conservation \cite{Noguchi:2007fi,Theers:2015fo}. 

Because SRD is an energy-conserving collision rule, a cell-level thermostat is
required to enforce isothermal conditions in nonequilibrium simulations.
Huang et al. \cite{Huang:2015fh} recently explored several thermostat schemes and found the Maxwell-Boltzmann
thermostat \cite{Huang:2010wt} to be most effective. Here, the current kinetic energy of each cell, $E_{\rm k}$, in the
reference frame of the cell center-of-mass velocity $\mathbf{u}$ is first determined,
\begin{equation}
E_{\rm k} = \frac{1}{2} \sum_{i=1}^{N_{\rm c}} m (\vv{v}_i - \vv{u})^2,\label{eq:thermo}
\end{equation}
where the sum is taken over the $N_{\rm c}$ particles in the cell.
Then, a random energy $\hat E_{\rm k}$ is drawn from the Maxwell-Boltzmann
distribution consistent with $N_{\rm c}$ at the desired temperature, and
the relative particle velocities are scaled by $\xi = \sqrt{\hat E_{\rm k} / E_{\rm k}}$.
This method generates average kinetic energies and fluctuations consistent with
the canonical ensemble \cite{Huang:2010wt,Huang:2015fh}.

Alternatively, the Andersen thermostat (AT) collision rule \cite{Allahyarov:2002hq} implicitly generates
isothermal conditions. In an AT collision, a random velocity $\delta \vv{v}_i$
consistent with the Gaussian distribution of velocities at the desired temperature is chosen
for each particle in a cell, and the velocities are updated by
\begin{equation}
\vv{v}_i(t+\Delta t) = \vv{u}(t) + \delta \vv{v}_i - \frac{1}{N_{\rm c}} \sum_{j=1}^{N_{\rm c}} \delta \vv{v}_j,\label{eq:at}
\end{equation}
where the sum is again taken over the particles in the cell. The last term
enforces linear-momentum conservation in the cell. Angular-momentum
conservation can also be enforced for the AT collision rule by applying an additional
constraint \cite{Noguchi:2007fi}.

Solute particles, such as the monomers of a polymer chain, can be coupled to the solvent
through the collision step \cite{Malevanets:2000tm}. Embedded particles propagate using standard molecular
dynamincs methods \cite{Allen:1991,Frenkel:2002chF} between MPCD collisions. Typically the MD timestep is much
shorter than the MPCD collision time in order to faithfully integrate the MD
equations of motion. The embedded particles are then binned into cells
with the solvent particles during the collisions. Care must be taken when
computing the center-of-mass velocity or kinetic energy of a cell to appropriately
weight the masses of the embedded particles, which are typically different from
the solvent.

There are many additional variations and features of the MPCD algorithm \cite{Gompper:2009is}, including the
modeling of fluid-solid boundaries \cite{Padding:2005ez,Poblete:2014gc,Bolintineanu:2014dj,Lamura:2001un,Whitmer:2010kp,Bolintineanu:2012ik},
nonideal \cite{Ihle:2006co} and multiphase \cite{Tuzel:2007fr} fluids, nonequilibrium flows \cite{Lamura:2001un,Allahyarov:2002hq},
viscoelasticity \cite{Tao:2008fk}, and deformable objects \cite{Noguchi:2005jz}.
A thorough treatment of these aspects of the algorithm is beyond the scope of the
current work. However, the open-source software presented here has been implemented
in a way that can be readily extended to support these features.

\section{Implementation\label{sec:implement}}
\begin{figure}[!htbp]
  \centering
  \includegraphics[width=3.5cm]{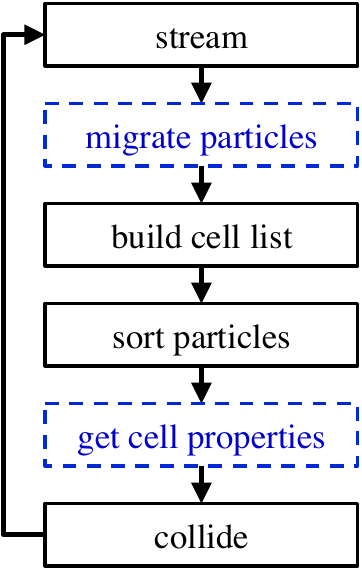}
  \caption{Flow diagram of a single step of the MPCD algorithm. Each block represents
           an independent step of the algorithm. Blocks with dashed outlines require
           nearest-neighbor MPI communication for multi-GPU simulations. Particle
           sorting is an optional step that is performed periodically.\label{fig:flow}}
\end{figure}

In this section, we present our implementation of the previously described MPCD
algorithms for GPUs. Throughout,
we will focus on optimizations for NVIDIA GPUs (Section~\ref{sec:hoomd}) and use
terminology from the CUDA programming model; however, many of the strategies
described here can be generalized to other multithreaded programming frameworks.
Figure~\ref{fig:flow} shows a flow diagram summarizing the steps of the MPCD
algorithm required to complete one time step of streaming and collision. Each
block represents an independent step in the algorithm, and blocks with
dashed outlines require communication with nearest-neighbor ranks in MPI. By
separating the steps of the algorithm in this way, we ensure a modular design
that can be readily extended.

An MPCD time step proceeds as follows. First, particles are streamed to their
new positions according to eq.~\ref{eq:stream}. Particles are then migrated onto the
appropriate rank (Section~\ref{sec:decomp}) and binned into cells (Section~\ref{sec:celllist}). Cell
properties such as the center-of-mass velocity are then computed (Section~\ref{sec:cellprop})
before finally the particles undergo a multiparticle collision according to,
e.g., eq.~\ref{eq:srd} or \ref{eq:at} (Sections~\ref{sec:srd} and~\ref{sec:at}). For performance reasons, the MPCD particle
data (Section~\ref{sec:pdata}) is optionally resorted periodically after the cell list is
constructed (Section~\ref{sec:sort}). Each of these points will be discussed in further
detail below.

\subsection{Framework\label{sec:hoomd}}
We chose to implement the MPCD algorithm in the HOOMD-blue simulation package \cite{Anderson:2008vg,Glaser:2015cu}.
HOOMD-blue is an open-source, general-purpose molecular dynamics code that was
designed for NVIDIA GPUs. It is primarily used to study soft matter and perform
coarse-grained simulations, and has a rich set of features for these purposes.
Unlike other MD software that uses the GPU as an accelerator to the CPU at
performance-critical steps, HOOMD-blue performs essentially all computation
directly on the GPU \cite{Anderson:2008vg}. This approach has the advantage that data structures can
be optimized for the GPU, and only limited amounts of data must be migrated
between the CPU and GPU. The efficient MD engine can be used
to simulate particles embedded within the MPCD solvent completely on the GPU.
(There is also a CPU-only version if a GPU is unavailable.) HOOMD-blue
additionally supports simulations on multiple CPUs and GPUs with MPI domain
decomposition and has been shown to scale to thousands of NVIDIA GPUs \cite{Glaser:2015cu}.
Moreover, it has a modular, object-oriented design and flexible Python user interface, which
make it straightforward to implement MPCD, as outlined in Figure~\ref{fig:flow},
alongside the existing MD engine and integrate it in users' simulation scripts.

\subsection{Particle data\label{sec:pdata}}
We considered two designs for the MPCD particle data structures: using HOOMD-blue's
existing MD particle data or creating a standalone container. The first
approach is more convenient for the programmer and is the design adopted in
the LAMMPS SRD package. However, considerably more information
is usually tracked for MD particles than is required for MPCD particles.
This can lead to massive memory waste that restricts the problem sizes that
can be studied. MPCD particles are also fundamentally different from
MD particles, and so may have different optimal data structures. For example, MPCD
particles nearly always have the same mass, and so it is more efficient to save
this value once for the fluid rather than per particle. Accordingly, we implemented
a standalone container for the MPCD particle data in HOOMD-blue.
\begin{figure}[!htbp]
  \centering
  \includegraphics[width=6cm]{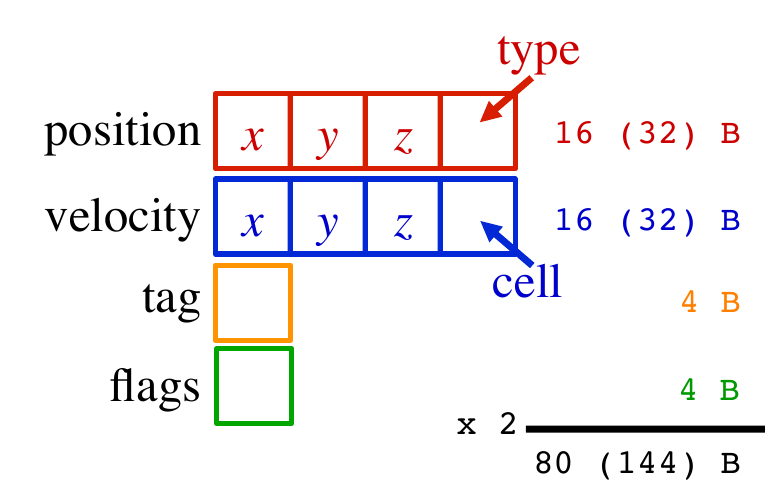}
  \caption{Particle data and memory requirements (in bytes) per MPCD particle.
           Each particle property is stored in a separate array.
           Position and velocity arrays are four-element vectors of 32-bit (or 64-bit) floating-point values,
           and tag and flag arrays are 32-bit integers. A duplicate is allocated for each
           array to facilitate sorting and particle removal, indicated by $\times 2$.\label{fig:pdata}}
\end{figure}

The particle data is stored as a structure of arrays. The entries in these arrays
for a single MPCD particle are schematically illustrated in Figure~\ref{fig:pdata}.
The position and velocity vectors of each MPCD particle are stored as
floating-point values. In order to promote coalesced memory accesses, the
particle positions and velocities are saved as arrays of four-element
floating-point vectors. The precision of these vectors can be selected
during compilation. The fourth elements of the vectors are used to store
additional data or as caches for data that may be commonly requested concurrently
with the position or velocity. Currently, an integer identifying the type of the
particle is stored in the position, and the current cell index of the particle
is cached in the velocity. For purposes of saving and manipulating configurations,
each particle is assigned a unique identifying integer tag. The particle tag is expected
to be rarely accessed during the simulation, and is used primarily for
manipulating particle data through the Python user interface. In MPI simulations,
an additional integer is allocated per particle to be used as a flag for
removing particles. A duplicate array is allocated for each property
in order to facilitate fast swapping of data as particles are sorted or 
removed. Hence, the particle data requires a maximum of 80 bytes
per particle for single-precision vectors and 144 bytes per particle for
double-precision vectors.

\subsection{Domain decomposition\label{sec:decomp}}
HOOMD-blue employs a standard spatial domain decomposition using MPI for
simulations on multiple GPUs \cite{Glaser:2015cu,Plimpton:1995fc}. The simulation box is subdivided by
planes along each of the box vectors, and one MPI rank is assigned to each
local simulation box. When multiple GPUs are available in a node, a two-tier
decomposition attempts to place spatially close
ranks within the same node. The MPCD particles are initially decomposed onto
each rank during initialization. They are subsequently migrated between
ranks when communication is requested by part of the algorithm. The communication
algorithm for the MPCD particles, which is similar to that employed for the
MD particles \cite{Glaser:2015cu}, is described below. All steps are performed on the GPU except where
explicitly noted otherwise.

Particles which have left the local simulation box are marked with an integer
consisting of bitwise flags denoting the directions each particle should be sent.
All marked particles are subsequently removed from the particle data and staged
into a send buffer, while the unmarked particles are compacted into the local
data arrays. After packing, the flags of particles in the send buffer are transformed into
destination MPI ranks, and the buffer is sorted by destination. Nonblocking MPI
calls then exchange particles with a maximum of 26 unique nearest neighbor ranks. If available,
a CUDA-aware MPI implementation \cite{Wang:2014id} could operate on buffers in device
memory directly, which would permit the MPI library to optimize these transfers
asynchronously and take advantage of GPUDirect technologies \cite{Shainer:2011jf,Potluri:2013by}. However,
in preliminary benchmarks, we found little performance gains from using
CUDA-aware MPI for this communication, and so instead the entire send buffer is
copied from the device to the host before the MPI calls are made. After the
communication is completed, received particles are wrapped back into their
destination simulation boxes and appended to the particle data arrays.

The CPU-only code employs a simpler communication pattern, as used in LAMMPS \cite{Plimpton:1995fc}, where
particles are exchanged with a maximum of 6 nearest neighbors along the cardinal directions in
three rounds of MPI calls. Particles received in each round are checked
to see if they require additional communication and, if so, are pushed into the
send buffer for the next round of communication. This approach increases the
message sizes in order to take advantage of the full network bandwidth
and decrease communication latency. However, on the GPU, additional kernel calls
and host-device data copies would be required in order to repack the send buffers
during each round of communication, leading to a serial bottleneck which
makes one round of communication with smaller messages more efficient.

\subsection{Cell list\label{sec:celllist}}
\begin{figure}[!htbp]
  \centering
  \includegraphics[width=6cm]{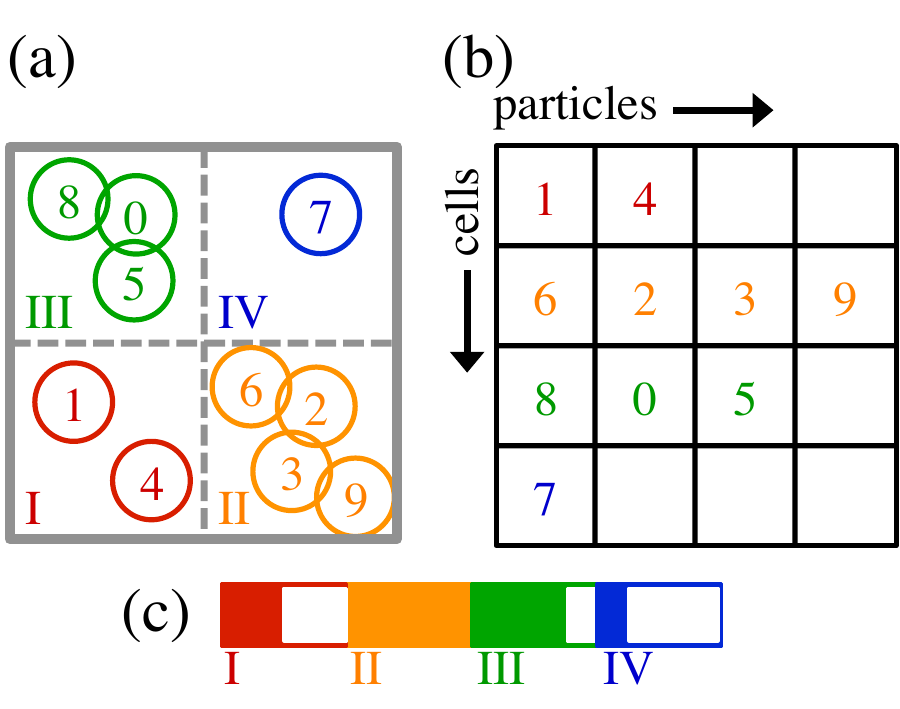}
  \caption{(a) Schematic illustration of particles binned into MPCD cell list.
           (b) Two-dimensional row-major layout of cell list with one row per cell
               and columns containing particles.
           (c) One-dimensional cell list memory with numbers indicating region
           for each cell and white space showing unused storage per cell.\label{fig:celllist}}
\end{figure}

Binning particles into local collision cells in order to compute average cell
properties like the center-of-mass velocity is central to the MPCD algorithm.
In principle, such properties can be computed by direct summation of particle
data. However, on the GPU, multiple atomic operations would be required to
perform the summation without a read-modify-write race condition, and such
operations are typically slow. Instead, it can be more efficient to first construct
a list of particles belonging to each cell (Figure~\ref{fig:celllist}a) \cite{Westphal:2014db},
and then perform the summation on particles within the cells. Moreover, a cell
list is more flexible for applying different collision rules
(see Section~\ref{sec:at}) and can be reused for efficient particle sorting (see Section~\ref{sec:sort}).

Westphal et al. proposed a sophisticated method for constructing a cell list
using hash tables and partial lists with shared memory staging \cite{Westphal:2014db}. In their
benchmarks, this method performed well on NVIDIA GPUs with the Fermi architecture,
but performance deteriorated for GPUs with the Kepler architecture that succeeded Fermi.
Since subsequent GPUs more closely resemble the Kepler architecture,
we require a different optimized solution. For example, the cell list could be
constructed by first assigning a cell index to each particle, and then sorting
the particle indexes using the cell index as a key. This approach has been
applied successfully in the HALMD package \cite{Colberg:2011de}, and it has the benefit
that the generated cell list can be stored using an array whose length is exactly
equal to the number of particles. However, sorting can be slow when the number of
key-value pairs is large, as is the case for MPCD. Additionally, an auxiliary
array is required to find the location of each cell's particles in compact
cell list storage, adding to the overhead of reading from the cell list.

In this work, we employ a method for generating the cell list based on limited use of atomic
operations that is also currently used in HOOMD-blue for the MD particles.
The cell list is stored as a two-dimensional matrix in row-major order \cite{Rapaport:2011el}. Each row
corresponds to a cell, and each column gives the index of a particle in that cell (Figure~\ref{fig:celllist}b).
The row-major ordering is beneficial for computing cell properties (see Section~\ref{sec:cellprop}).
Because of the regular data layout, entries in the cell list can be straightforwardly
accessed without an auxiliary array. However, enough memory must be allocated
so that each cell can hold the maximum number of particles in any cell. This layout leads
to some memory waste, particularly if one cell has many more particles than another (Figure~\ref{fig:celllist}c).
The number of MPCD particles in a cell should follow a Poisson distribution \cite{Gompper:2009is,Huang:2010wt},
and we found that the typical cell list memory was only roughly 4 times larger than a
perfectly compact array in our benchmarks (see Section~\ref{sec:perf}). This overhead is
comparable to the total memory requirements to build a cell list using an efficient
parallel radix sort algorithm \cite{Merrill:2011vy,cubweb}.

The cell list is built as follows. Each particle is first binned into a cell (Figure~\ref{fig:celllist}a).
In order to support random grid shifting, particle positions are optionally
shifted by a vector, subject to periodic boundary conditions, during binning.
A particle is then inserted into the row of its cell using one atomic operation
to determine the column. The cell index is also cached into the fourth element
of the particle velocity vector for use in subsequent steps of the MPCD algorithm.
The particle is not inserted if the determined column index would exceed the
allocated size of the cell list. However, in this case, the maximum number of particles
that should be inserted into any cell is still implicitly tracked, and the
cell list is subsequently reallocated and recomputed. Reallocation
occurs infrequently and primarily during the first few times the cell list is built.

\begin{figure}[!htbp]
  \centering
  \includegraphics[width=5cm]{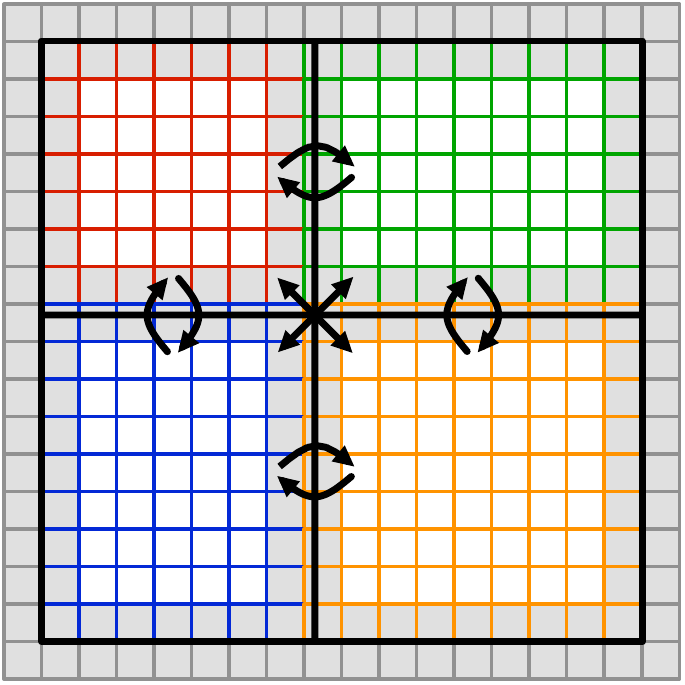}
  \caption{Schematic of MPCD cells with domain decomposition. The global and local
           simulation box boundaries are indicated by solid black lines. Cells
           requiring communication, including ghost cells outside global simulation
           box, are shaded grey. Arrows schematically illustrate the point-to-point
           MPI communication pattern with neighboring ranks used in
           both particle migration (Section~\ref{sec:decomp}) and cell property
           calculation (Section~\ref{sec:cellprop}) for the GPU.
           \label{fig:decomp}}
\end{figure}

In MPI simulations, cells may overlap the boundaries of domains. This
is further complicated by grid shifting, since particles may be shifted outside
the local simulation box. To accommodate these boundaries, we introduce a layer of
ghost cells around the domain (Figure~\ref{fig:decomp}). We compute the minimum number of cells required
along each dimension so that a particle shifted from the local simulation box by
$\pm a/2$ will still be binned into a cell. Because of this assumption, particles
must be migrated to lie within their local simulation boxes before the cell list
is built, effectively requiring that particle migration occurs before every MPCD
collision. In order to reduce the frequency of particle migration, it has been
suggested to use additional ghost cells as a buffer so that particles can diffuse farther
off rank \cite{Sutmann:2010jr}. We found that using additional ghost cells was not effective for the
GPU because increased cell communication posed a bigger bottleneck.

\subsection{Cell properties\label{sec:cellprop}}
After the cell list has been built, the cell properties can be computed by iterating
over particles in each cell. The main quantities of interest are the center-of-mass
velocity and the kinetic energy. The temperature in the cell, which is defined
by the kinetic energy relative to the center-of-mass velocity (see eq.~\ref{eq:thermo}), can be determined from
these properties. The simplest GPU implementation of this calculation would use
one thread per cell to loop over the member particles and accumulate the momentum
and kinetic energy \cite{Westphal:2014db}. However,
this approach has limited parallelism, especially when scaling to multiple GPUs,
because the number of cells may be too small to fully occupy a GPU.

Instead, we compute the cell properties using an array of $w$ threads to process the
particles in a cell. Using multiple threads increases parallelism and also
promotes coalesced reads from the row-major ordered cell list. Each thread reads
from the cell list in a strided fashion to compute the total momentum, mass, and
kinetic energy of the particles it has accessed. The number of threads $w$ is
restricted to be a power of two less than the size of a CUDA warp so that these
quantities can then be reduced using shuffle instructions available on Kepler
and newer NVIDIA GPUs that allow threads within
a warp to efficiently read from each other's registers. The first thread within
each array then computes the center-of-mass velocity from the momentum, determines
the temperature, and writes the final result. The optimum value of $w$ depends
on the MPCD particle density, the volume of the simulation box, and the specific
GPU hardware and is determined using runtime autotuning \cite{Glaser:2015cu}.

It is essential to accurately compute the cell center-of-mass velocity because
the MPCD collisions are performed relative to this quantity, and inaccuracies
can lead to momentum drift and poor energy conservation. Accordingly, all
reductions are performed with double-precision floating-point values even if the
particle velocities are stored in single precision. Alternatively, we could perform
the operations using double-single floating-point arithmetic \cite{Colberg:2011de} or fixed-point arithmetic \cite{LeGrand:2013wc}.
These strategies replace 64-bit floating-point arithmetic with multiple equivalent 32-bit floating-point operations
or 64-bit integer fixed-point arithmetic, respectively, which can be faster for certain GPU architectures
due to differences in instruction throughput. However, we found that the primary
bottleneck in the cell property calculation was actually the memory bandwidth
rather than arithmetic operations, and so chose to simply use
double-precision floating-point operations for reducing the cell properties
(see Section~\ref{sec:prec}).

In MPI simulations, communication between ranks is required in order to compute
properties in cells overlapping the boundaries. This communication pattern is
regular provided that the domain decomposition does not change during the simulation.
We adopt a similar communication pattern as for the particle migration on the GPU,
and employ nonblocking MPI calls to a maximum of 26 nearest neighbors for both
CPU and GPU code pathways. All buffers are packed and unpacked on the GPU if
available, and only the minimum required amount of data is communicated.
Still, the amount of data that must be transferred from the device to host is
large, leading to a significant latency. To mask this latency, we divide
the computation of cell quantities into two stages. First, we compute the
properties for ``outer'' cells requiring communication (shaded cells in Figure~\ref{fig:decomp}). We then begin the MPI
communication, and calculate properties of cells lying fully within the local
simulation box (white cells in Figure~\ref{fig:decomp}). Last, we finalize communication from the outer cells and finish
reducing their properties. We found this strategy to be much more efficient than performing
three rounds of communication with fewer, larger messages because it (1) does not
lead to serial bottlenecks of host-device data migration and (2) allows computation
to overlap with network communication.

The total momentum, energy, and temperature of the system may be required for
measurement purposes. These quantites can be obtained by summation of the cell
properties, which are reduced using a parallel device-wide reduction scheme.
In MPI simulations, we ensure that cells overlapping the boundary are only
included in this sum by one of the ranks. The net properties are then reduced
across all MPI ranks with a collective communication call.

\subsection{Stochastic rotation dynamics\label{sec:srd}}
Once the MPCD cell list and cell properties have been computed, it is straightforward
to apply the SRD collision rule to the particle velocities. The most important
technical challenge to address is how to randomly draw and store the rotation matrix
for each cell. The rotation vector must be picked randomly and uniformly from the
surface of the unit sphere. We employ the cylindrical projection method for picking
points, which requires drawing only two uniform random numbers. This method is
particularly efficient on the GPU compared to rejection sampling methods because
there is no branch divergence. The rotation axis for each cell is stored in
double precision to ensure numerical accuracy of the rotation.

We generate the uniform random numbers using a cryptographic-hash approach to create
unique independent microstreams of random numbers \cite{Phillips:2011td}. This approach has proven
to be highly useful for massively parallel computing because it allows independent
threads to generate random numbers without needing to access or advance a shared
state \cite{Salmon:2011eg}. In this work, we employ the Saru random number generator used throughout
HOOMD-blue, although other generators
could be straightforwardly substituted. Saru takes three seeds to its hash to
initialize a compact state \cite{Phillips:2011td}. To generate random vectors for the cell, we feed
three seeds: (1) the unique global index of the cell (within the entire simulation
box), (2) the current simulation timestep, and (3) a user-supplied seed. This
choice of seeding ensures a unique stream (1) for each cell, (2) at each time
step, and (3) between simulations. Moreover, because the global cell index is
used, the chosen rotation vector for any given cell can be reproduced at any
given time step on any rank, regardless of the previous history of drawing numbers.
This is particularly useful for MPI simulations because it eliminates the need
to communicate random rotation vectors between ranks, significantly decreasing latency.

Particle velocity rotation simply proceeds as outlined in Algorithm~\ref{alg:srd}. Each particle velocity is loaded
into memory, including its cell index, which was stored as the fourth element
of its velocity (line 1). The average velocity (line 2) and rotation axis (line 3) for the cell are read
from memory, and the rotation is applied to the relative particle velocity (lines 6--7).
The center-of-mass velocity is added back to the rotated relative velocity (line 11), and
the updated velocity is stored (line 12). We found that, regardless of the precision of the
particle velocities, it was necessary to perform the rotations in double precision
in order to obtain good momentum conservation. We accordingly upcast the particle
velocities to double precision before performing any steps in the rotation (line 4).
The precision of the rotated velocity vector is then downcast for storage (line 12). We will
discuss the ramifications of this type conversion for accuracy of the simulations
in Section~\ref{sec:prec}.

\begin{algorithm}
\caption{SRD collision rule.} \label{alg:srd}
\begin{algorithmic}[1]
\ParallelForEach{particle $i$}
    \State $\{\mathbf{v}_i,c\} \gets$ particle velocity and cell
    \State $\mathbf{u} \gets$ center-of-mass velocity for $c$
    \State $\mathbf{R} \gets$ rotation vector for $c$
    \item[]
    \State Cast $\mathbf{v}_i$ to double precision.
    \State $\Delta \mathbf{v}_i \gets \mathbf{v}_i - \mathbf{u}$
    \State Rotate $\Delta \mathbf{v}_i$ by angle $\alpha$ around $\mathbf{R}$.
    \If{thermostat enabled}
        \State $\xi \gets$ scale factor for $c$
        \State $\Delta \mathbf{v}_i \gets \xi \Delta\mathbf{v}_i$
    \EndIf
    \State $\mathbf{v}_i \gets \mathbf{u} + \Delta \mathbf{v}_i$
    \State Store $\mathbf{v}_i$ in native precision.

\EndParallelForEach
\end{algorithmic}
\end{algorithm}

As discussed in Section~\ref{sec:alg}, the SRD collision rule optionally applies a cell-level
thermostat. This step requires first computing the rescaling factor $\xi$ per cell according
to eq.~\ref{eq:thermo} and then applying this factor to the relative particle velocities. If
thermostatting is enabled, we compute $\xi$ for each cell at the
same time that we draw the rotation vectors. A random kinetic energy is drawn for
each cell from a $\Gamma$ distribution using Marsaglia's efficient rejection
sampling method \cite{Marsaglia:2000kp}. This method requires drawing a Gaussian random variable,
which we generate by the Box-Muller transformation \cite{BOX:1958wm} of two uniform variables
to avoid additional branch divergence \cite{Riesinger:2014wi}. Cells having fewer than two particles are
assigned a scale factor of $\xi=1$ because the temperature is not defined in these cases.
Velocity rescaling is then applied to the relative velocities after rotation
but before they are shifted by the center-of-mass velocity and stored (lines 8--10 in Algorithm~\ref{alg:srd}).

Solute particles are coupled to the MPCD solvent during the collision step. This
necessitates including the solute particles in all stages of the algorithm, including
cell list construction, cell property calculation, and then the velocity rotation.
In HOOMD-blue, the coupled particles are treated as a subset (group) of MD particles.
The solute particles evolve according to the standard MD equations of motion
between collisions using the velocity Verlet algorithm \cite{Allen:1991,Frenkel:2002chF}.
Since the solute is coupled only during the collision step, the MPCD particles are
only required to be streamed every MPCD collision time $\Delta t$ and not every MD time step.

\subsection{Andersen thermostat\label{sec:at}}
The AT collision rule requires the generation of a set of random velocities per
MPCD particle and the subsequent reduction of these velocities within each cell
in order to apply eq.~\ref{eq:at}. We first draw random velocities for each particle from
a Gaussian distribution using the Box-Muller transformation. The uniform random
values are generated using Saru with the timestep, particle tag, and a
user-supplied seed. These velocities are then summed using the methods described
in Section~\ref{sec:cellprop}. The particle velocities are then updated by
adding the randomly drawn velocity for each particle to the cell center-of-mass
velocity and subtracting the summed random contribution. The velocity update step
is simpler than the SRD algorithm (Algorithm~\ref{alg:srd}) because only simple
summation is required and the AT collision rule implicitly thermostats the solvent.
However, we found that the AT collision rule performed slower than
SRD overall for two reasons: (1) more random numbers must be drawn (per-particle
rather than per-cell), and (2) an additional reduction is performed for each cell.
This second step especially incurs a performance penalty in MPI simulations because
additional communication is required compared to SRD. To help mask some of this
latency, we overlap the process of drawing the random velocities with communication
during the initial calculation of the center-of-mass velocities for each cell.

\subsection{Particle sorting\label{sec:sort}}
Many steps of the MPCD algorithm involve processing particles within one cell.
Performance of these steps can be improved by first reordering the particle data
into cell order \cite{Westphal:2014db}, which improves data locality. HOOMD-blue uses a three-dimensional
Hilbert curve to sort the MD particles \cite{Anderson:2008vg}, which is beneficial for evaluating pair
forces between particles. The MPCD particles can be sorted more simply into the
order they reside in the cell list since this strategy gives optimal ordering for
cell property calculation during the first MPCD collision after sorting. Random
grid shifting, particle diffusion, and insertion order of particles into
the cell list will decrease this ordering on subsequent collisions, but
particles still retain some locality. Using the cell list to sort the particles
is particularly convenient because it does not require significant additional
calculations, and sorting can be injected into the usual MPCD algorithm
(see Figure~\ref{fig:flow}). The ability to sort the MPCD and MD particles independently is
another advantage of the self-contained MPCD particle data structure described
in Section~\ref{sec:pdata}.

The particles are sorted as follows. First, the cell list is constructed
as described in Section~\ref{sec:celllist}. If sorting should occur
at the current MPCD step, the two-dimensional cell list array (Figure~\ref{fig:celllist}b)
is compacted into a one-dimensional list of MPCD solvent particle indexes. The
compaction step is necessary in order to remove any empty entries from cells
with fewer particles than the maximum allocation per cell (Figure~\ref{fig:celllist}c).
Embedded solute particle indexes are also removed to preserve the Hilbert-curve ordering
already applied independently to the MD particles. The embedded particles usually
comprise only a small fraction of particles in the cells, and so their
ordering should have negligible impact on performance. The MPCD particle data
is then sorted according to the compacted list, and the indexes in the cell
list are updated to the new ordering.

Although particle sorting improves performance of other steps of the MPCD algorithm,
there is a significant cost associated with data movement during the sort,
especially on the GPU. There is accordingly an optimum frequency of sorting that
balances the improved performance from data locality with this added cost. The
optimum will depend on the specific properties of the MPCD fluid, the number of particles per GPU,
and also the GPU architecture. We therefore advocate tuning the sorting frequency
with a series of short simulations to achieve maximum performance.

\section{Performance\label{sec:perf}}
Performance was benchmarked using the SRD collision rule with typical simulation parameters.
MPCD particles having unit mass $m$ were randomly placed into a cubic simulation box of
edge length $L$ at number density $\rho = 10/a^3$. Here, the MPCD cell size $a$ defines
the unit of length. The particle velocities were drawn randomly from the Gaussian
distribution consistent with temperature $T = \varepsilon / k_{\rm B}$, where
$\varepsilon$ is the unit of energy.
The SRD rotation angle was $\alpha = 130^\circ$, and the time between collisions
was $\Delta t = 0.1\,\tau$, where $\tau = \sqrt{m a^2/\varepsilon}$ is the unit
of time. Random grid shifting was applied to ensure Galilean
invariance. With this choice of parameters, the solvent viscosity was
$8.7\,\varepsilon \tau/a^3$ and the estimated Schmidt number was 14 \cite{Huang:2010ed}.
These values are consistent with a liquid-like solvent \cite{Ripoll:2004jn,Ripoll:2005ev}. The MPCD particle data
was sorted every 25 collisions.

We performed four benchmarks of our MPCD implementation. We first tested the accuracy and
performance of the mixed-precision model (Section~\ref{sec:prec}) for this SRD fluid.
We then assessed the weak and strong scaling efficiency of our implementation (Section~\ref{sec:scaling}).
Finally, we performed a research-relevant benchmark of polymer chains in solution (Section~\ref{sec:polymer}).
These benchmarks are described in detail next and summarized in Table~\ref{tab:bench}.

\begin{table}[h]
\caption{Summary of benchmarks. $L$ is the edge length of the simulation box, and $N$ is the number of MPCD particles.
For the weak scaling benchmark, $L$ refers to the size of a cubic box per node.
The MPCD parameters for all benchmarks are density $\rho = 10/a^3$,
temperature $T = \varepsilon/k_{\rm B}$, SRD rotation angle $\alpha = 130^\circ$, and collision time
$\Delta t = 0.1\,\tau$.\label{tab:bench}}
\centering
\begin{tabular}{lcccc}
benchmark & $L$ & $N$ & simulated time & hardware \\
\hline
\multirow{2}{*}{precision (Section~\ref{sec:prec})} & $50\,a$ & 1.25 million & \multirow{2}{*}{$10^5\,\tau$} & Tesla P100 / \\
          & $100\,a$ & 10 million & & GeForce GTX 1080\\
weak scaling (Section~\ref{sec:scaling}) & $50\,a$ / node & 1.25 million / node & $500\,\tau$ & Blue Waters (Tesla K20x)\\
strong scaling (Section~\ref{sec:scaling}) & $400\,a$ & 640 million & $500\,\tau$ & Blue Waters (Tesla K20x) \\
polymer (Section~\ref{sec:polymer}) & $50\,a$ & 1.25 million & $2000\,\tau$ & Tesla P100 \\
\hline
\end{tabular}
\end{table}

\subsection{Precision\label{sec:prec}}
HOOMD-blue's MD codes can be compiled to use single-precision or double-precision
floating-point values for its data structures and calculations. Single precision
can give a sizable performance increase compared to double precision due to
reduced data size and higher 32-bit floating-point arithmetic instruction throughput.
Our MPCD implementation is often bottlenecked by accessing particle data in global
memory, and so MPCD should similarly benefit from using single-precision
floating-point values when possible. In what we will refer to as the mixed-precision model
for our implementation, particle positions and velocities are
stored in single precision instead of double precision (Section~\ref{sec:pdata}), and single-precision
floating-point operations are used when possible in the algorithm. However, some key steps always
require double-precision values and arithmetic for numerical accuracy
(Sections~\ref{sec:cellprop} and~\ref{sec:srd}), prohibiting the exclusive use of single-precision
floating-point operations throughout. In these steps,
the single-precision particle data is upcast to double precision, and intermediate calculations are performed using
double-precision values before downcasting the result to single precision. Here, we
study the numerical accuracy and performance of the SRD algorithm for this mixed-precision model
compared to a fully double-precision model.

\begin{figure}[!htbp]
  \centering
  \includegraphics{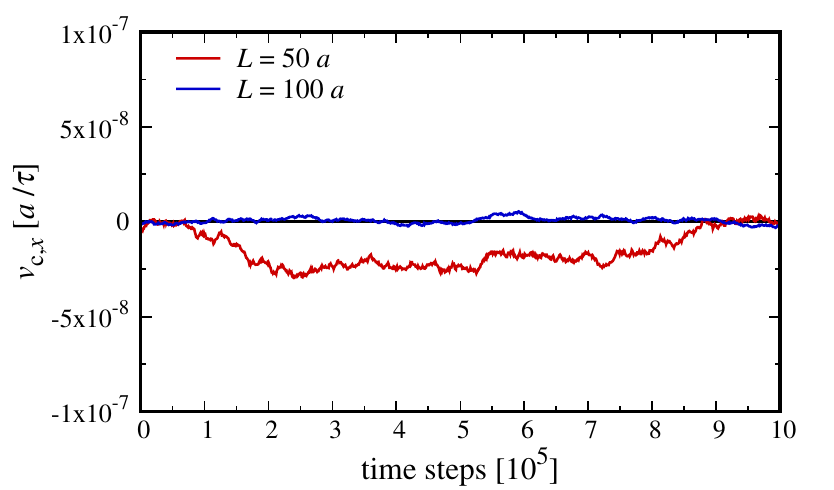}
  \caption{Center-of-mass velocity of benchmark SRD fluid along $x$ direction, $v_{{\rm c},x}$, in mixed-precision
           model over $10^6$ MPCD time steps for box sizes $L = 50\,a$ and $100\,a$.\label{fig:drift}}
\end{figure}

We measured the $x$-component of the center-of-mass velocity $v_{{\rm c},x}$ of the benchmark SRD fluid
for box sizes $L=50\,a$ and $L=100\,a$ over $10^6$ MPCD time steps. (Comparable results were obtained for
the $y$ and $z$ components.) Random initialization of the particle
velocities imparted a nonzero initial value, which we aimed to remove by distributing
a constant correction to each particle, as is the typical strategy \cite{Allen:1991,Frenkel:2002chF}.
For the double-precision model, $v_{{\rm c},x}$ was essentially zero for all times considered after the initial
correction was applied ($|v_{{\rm c},x}| \lesssim 10^{-16}\,ma/\tau$). For the mixed-precision model,
however, this process still left a small nonzero value due to loss of precision when the velocities
were stored in single precision after the correction. The center-of-mass velocity then fluctuated around zero during the
simulation (Figure~\ref{fig:drift}). Importantly, we did not observe any significant systematic drift over
the simulation times considered, corresponding to good global momentum conservation.
The absolute value of $v_{{\rm c},x}$ remained small and bounded, and decreased for larger numbers of
particles (larger $L$). It appeared that $v_{{\rm c},x}$ for the mixed-precision model was essentially
a random variable due to the digits that were lost when casting the precision of the particle velocities for rotation
since a single-precision floating-point value typically contains only 7 or 8 significant digits \cite{ieee754}.
Comparable values of $v_{{\rm c},x}$ are obtained in MD simulations with good momentum conservation \cite{Colberg:2011de}.
We accordingly judged that the accuracy of the mixed-precision model was acceptable and
that mixed precision could be used provided that it improved performance.

\begin{figure}[!htbp]
  \centering
  \includegraphics{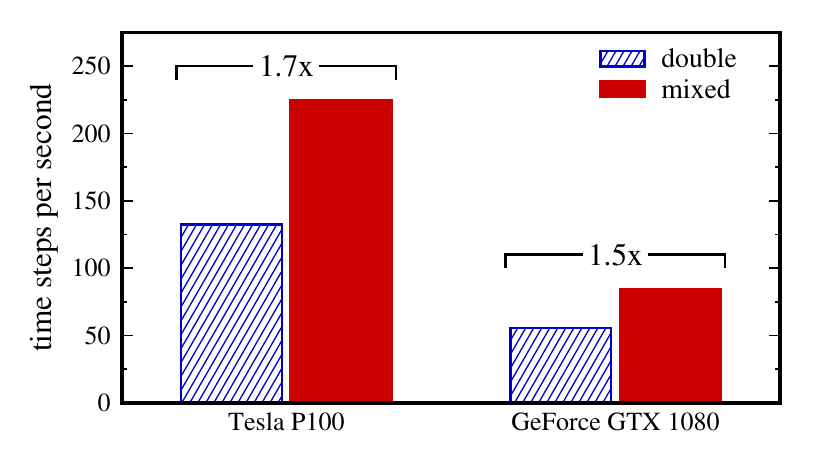}
  \caption{Time steps per second for SRD collision rule in double precision
           and mixed precision on NVIDIA Tesla P100 and GeForce GTX 1080 GPUs
           for the $L=100\,a$ simulation box.\label{fig:precision}}
\end{figure}

We tested the performance of the double- and mixed-precision models on two recent
NVIDIA GPUs, Tesla P100 and GeForce GTX 1080, using CUDA 8.0. The
average performance in time steps per second (larger is faster) is shown in Figure~\ref{fig:precision} for the
$L=100\,a$ box. For the Tesla P100, the mixed-precision model was
1.7x faster than the double-precision model. This difference is close to the
maximum 2x speedup that could be obtained from the difference in peak theoretical
floating-point performance in single and double precision. For the GeForce GTX 1080,
the speedup from the mixed-precision model was slightly smaller at 1.5x. Our
benchmarks indicate performance is mostly limited by data accesses, which
is perhaps unsurprising for the MPCD algorithm where many key steps involve only
little computation. The MPCD algorithm benefits accordingly from the
mixed-precision model, which reduces the size of the particle data while
delivering reasonable accuracy and performance.

\subsection{Scaling\label{sec:scaling}}
Efficiently utilizing multiple GPUs is a requirement for studying
problems too large to reside in the memory of a single GPU and also
decreases the time required to complete a simulation. We
performed scaling benchmarks of our MPCD implementation on Blue Waters, hosted
by the National Center for Supercomputing Applications at the University
of Illinois at Urbana-Champaign. HOOMD-blue was compiled using the GNU Compiler
Collection (version 4.9.3) and CUDA 7.5 with the double-precision model to
provide a performance baseline. The GPU code was tested using Blue Waters's Cray XK7 nodes, each of which contains
one NVIDIA Tesla K20x GPU and one AMD Interlagos 6276 CPU with 8
floating-point cores. Previous benchmarks of HOOMD-blue's MD code found no
benefit of multiplexing multiple MPI processes onto the same GPU, and we expect
the same to hold true for MPCD. Accordingly, one MPI rank was assigned per XK7 node. 
The CPU code was tested using Blue Waters's Cray XE6 nodes with 16 MPI ranks per node.
Each XE6 node has two AMD Interlagos 6276 CPUs (16 total floating-point cores).
We chose to compare the CPU and GPU code using these configurations to obtain a
practical test of the maximum performance that could be achieved per node hour.

In order to test scaling of the MPCD code to problems with large length scales,
we performed a weak scaling test with a constant number of particles per node.
A cubic box with edge length $50\,a$ was assigned per node so that there were
1,250,000 particles per node on average. We first performed 2,000 SRD time steps
in order to allow all runtime autotuners to determine their optimal parameters.
These parameters were then fixed, and the performance was measured over 5,000
SRD time steps with HOOMD-blue's internal profiler enabled. We found that
overhead from profiling incurred a small performance penalty of roughly 5\% for
the GPU in this test. Each benchmark was repeated three times to determine the average performance,
and the 95\% confidence interval was estimated from these measurements.

\begin{figure}[!htbp]
  \centering
  \includegraphics{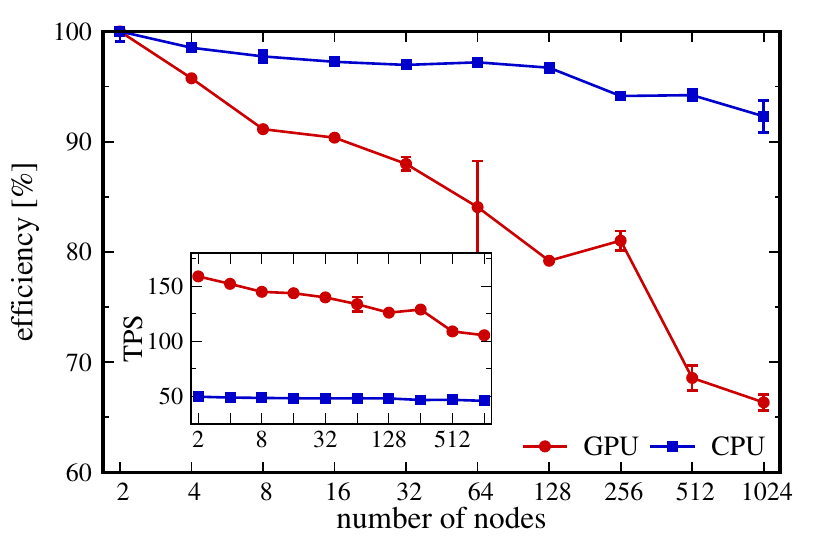}
  \caption{Weak scaling efficiency for benchmark SRD fluid on NCSA Blue Waters.
           One cubic simulation box with $L = 50\,a$ was replicated per node so
           that there were 1,250,000 particles per node on average. GPU benchmarks
           (circles) were performed using 1 MPI rank (1 NVIDIA Tesla K20x GPU) per XK7 node, while CPU
           benchmarks (squares) used 16 MPI ranks (16 floating-point cores) per XE6 node. The inset shows
           the performance in time steps per second (TPS).\label{fig:weak}}
\end{figure}

Both the CPU and GPU codes gave excellent weak scaling performance up to 1,024
nodes, as shown in Figure~\ref{fig:weak}. At this largest node count, there
were over one billion MPCD solvent particles in the simulation box. The CPU weak
scaling efficiency relative to two nodes was greater than 90\% for all node
counts tested, while the GPU scaling efficiency was greater than 65\%. Most of
the loss of efficiency in the GPU code came from an increase in the time required
to compute cell properties, which requires the most communication in our
implementation. There was also a large initial drop in efficiency for the GPU
code as the domain decomposition was increased from one dimension (at 2 nodes)
to three dimensions (at 8 nodes). This drop was expected because the amount of
communication per node increased. The CPU code did not show this initial drop
because there were 16 MPI ranks per node, and so the simulation box
was three-dimensionally decomposed for all node counts. The inset to Figure~\ref{fig:weak}
shows the absolute performance of the two codes in time steps per second. The
GPU code was roughly 3 times faster than the CPU code on 2 nodes and 2 times
faster on 1,024 nodes. For reference, the theoretical peak double-precision performance
of the K20x is roughly 4 times the XE6 node.

We then tested the strong scaling performance of the MPCD code for a large simulation
box with $L = 400\,a$, which correponded to $640$ million solvent particles.
Efficient strong scaling is required to fully utilize computational
resources and decrease the time to solution. As for the weak scaling tests, we
measured the time required to complete 5,000 SRD time steps with internal profiling
enabled after runtime autotuning was completed. The performance overhead
of this profiling was again roughly 5\% on the GPU for this test, and was most significant
at the largest node counts. We measured the scaling from 32
nodes up to 1,024 nodes. Although the CPU code could be run on smaller node counts,
the GPU code required a minimum of 32 nodes due to the limited capacity of
the GPU memory (6 GB). Each benchmark was repeated three times.

\begin{figure}[!htbp]
  \centering
  \includegraphics{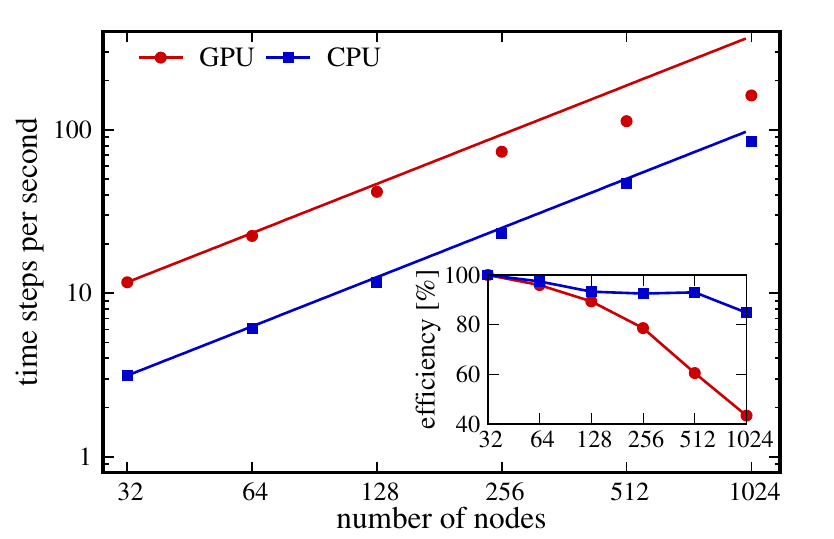}
  \caption{Strong scaling performance for benchmark SRD fluid in a cubic simulation
           box with $L = 400\,a$ on NCSA Blue Waters. The total number of particles was
           $N = 640,000,000$. GPU benchmarks (circles) were performed using 1 MPI rank (1 NVIDIA Tesla K20x GPU) per XK7 node, while CPU
           benchmarks (squares) used 16 MPI ranks (16 floating-point cores) per XE6 node. Solid lines indicate
           ideal scaling, and the inset shows the strong scaling efficiency.\label{fig:strong}}
\end{figure}

\begin{figure}[!htbp]
  \centering
  \includegraphics{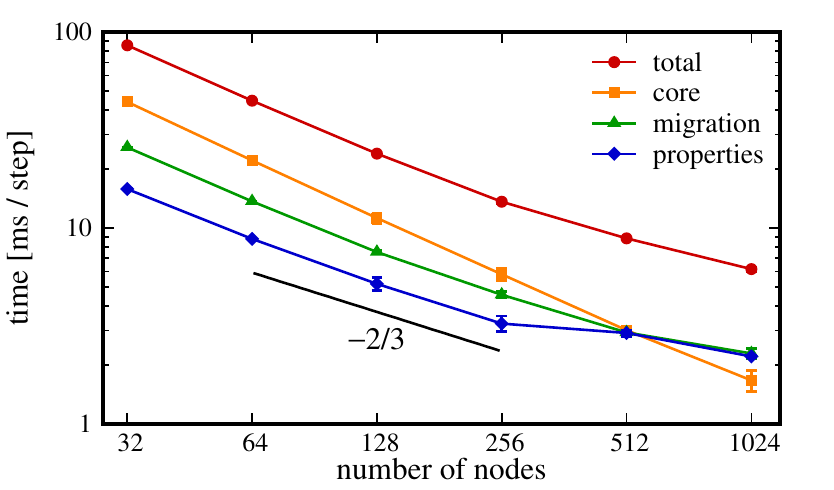}
  \caption{Profile of GPU strong scaling for SRD fluid on NCSA Blue Waters (see Figure~\ref{fig:strong}). Required
           time per step is shown for the total simulation (circles), core SRD methods (squares),
           particle migration (triangles), and computing cell properties (diamonds).
           Solid black line indicates ideal scaling of the cell property communication.\label{fig:prof}}
\end{figure}

Figure~\ref{fig:strong} shows the strong scaling performance and efficiency (inset)
of both the CPU and GPU codes. The CPU code exhibits excellent strong scaling,
with over 80\% efficiency on 1,024 nodes (16384 MPI ranks). The GPU code exhibits
similarly good scaling, with some efficiency lost at high node counts. In order
to investigate this loss of efficiency further, we separated the time required
per step into three main components: particle migration, cell property calculation,
and all other steps (including streaming and collision) that require no communication.
Figure~\ref{fig:prof} shows these components for various node counts. It is
clear that the communication steps, especially cell property calculation, limit
the scaling efficiency at high node counts.

At small node counts, the cell property calculation is in principle dominated by the
inner cells, which can effectively mask most of the latency from communication.
At larger node counts, however, the latency associated
with both host-device data migration as well as the MPI communication itself
become significant. Theoretically, it is expected that the time for communication of cell
properties should scale with $1/P^{2/3}$ for $P$ MPI ranks due to the
surface-to-volume ratio of the domain decomposition \cite{Sutmann:2010jr}. Initially, we did indeed observe
this scaling as indicated in Figure~\ref{fig:prof}, but it is clear that
for $P > 256$ much weaker scaling was obtained. More detailed
profiling revealed that this bottleneck was due to host-device data
migration. This latency might be reduced using an MPI library that operates on
device buffers directly since the library can optimize copies asynchronously
to pipeline messages or take advantage of GPUDirect RDMA technologies \cite{Shainer:2011jf,Potluri:2013by,Wang:2014id}.
Unfortunately, we found essentially no performance benefits from these optimizations on
Blue Waters due to limited support for GPUDirect RDMA.

Overall, the GPU code is faster than the CPU code by a factor between 2 and 3.
As a point of reference, we note that our CPU code also outperforms a comparable
benchmark with LAMMPS SRD (31 March 2017 release) \cite{Plimpton:1995fc,Petersen:2010cg} by nearly a factor of 2 when
running on 128 XE6 nodes on Blue Waters. (There are some minor SRD parameter differences
in the LAMMPS benchmark due to limitations of that implementation.) Our weak and
strong scaling benchmarks clearly demonstrate the feasibility of using GPUs to
efficiently perform and accelerate MPCD simulations at massive scale.

\subsection{Polymer solution benchmark\label{sec:polymer}}
In addition to the pure SRD fluid benchmarks, we also performed a more complex
simulation of a solution of polymers embedded in the benchmark SRD
solvent on an NVIDIA Tesla P100 GPU using the double-precision model and CUDA 8.0.
The polymers were represented by a bead-spring model as described in \cite{Huang:2010ed}.
Each polymer chain consisted of 50 beads, and the number of chains was varied
from 16 to 1,024, giving between 800 and 51,200 total monomers in the
simulation box with $L=50\,a$. The equations of motion for the polymers
were integrated between collisions with the solvent using the velocity Verlet
algorithm with a time step of $0.002\,\tau$. (An MPCD collision occurred every
50 MD timesteps.) The MD neighbor list used to compute nonbonded pair forces
had a buffer radius of $0.4\,a$. The MPCD solvent particles were streamed every
50 MD timesteps because their updated positions were only required for collisions.
We performed a short run of $200\,\tau$ to determine the optimal runtime kernel
launch parameters and then profiled performance for 2,000$\,\tau$.

In order to measure the contribution of the MPCD solvent to the simulation time,
we performed simple Langevin dynamics simulations \cite{Phillips:2011td} of the polymers as a baseline.
To obtain comparable long-time polymer dynamics, we adjusted the monomer
friction coefficient \cite{Howard:2017} to give a long-time polymer diffusion coefficient consistent
with the value predicted by the Zimm model for dilute polymer solutions \cite{Zimm:1956ue,Doi,Dunweg:1998va}. We
emphasize that this baseline does not include any hydrodynamic interactions,
which is a clear advantage of the MPCD simulations, and simply serves as a point
of comparison of the contributions of MD and MPCD to the overall simulation time.

\begin{figure}[!htbp]
  \centering
  \includegraphics{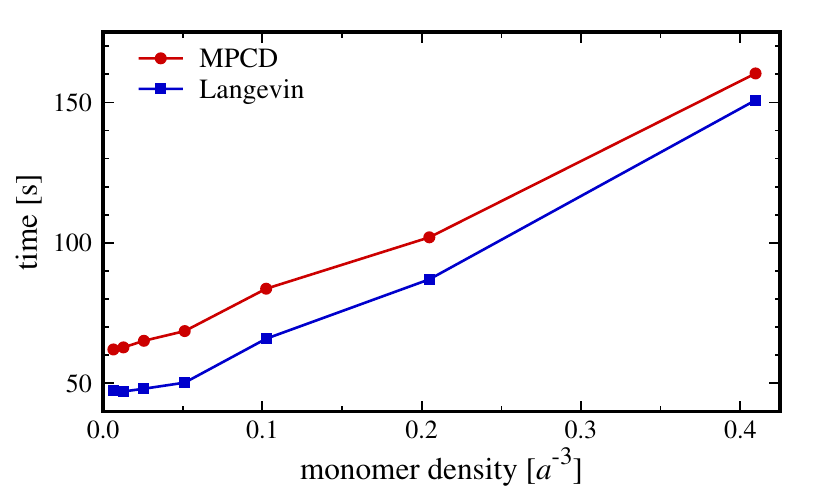}
  \caption{Required time to simulate polymer solutions at various concentrations
           for 2,000$\,\tau$ using MPCD (circles) and Langevin dynamics (squares)
           on an NVIDIA Tesla P100 GPU.\label{fig:poly}}
\end{figure}

The total required simulation time increased nearly linearly with the monomer
density, shown in Figure~\ref{fig:poly}. The density spanned from the dilute
to semidilute polymer concentration regimes \cite{Huang:2010ed}. Initially, the Langevin dynamics
simulation time did not increase much with density due to limited occupancy of the GPU
for small numbers of MD particles. The required time for the MPCD simulations closely
tracked the Langevin dynamics simulations, indicating that the performance of
the MPCD algorithm itself did not depend significantly on the polymer concentration.
This trend is reasonable because the embedded particles comprise only a small
fraction of the particles in an MPCD cell.
The MPCD simulations were at most 1.4x slower than the Langevin dynamics
simulations, which is only a modest cost to incorporate hydrodynamic interactions.
Even though there were many more solvent particles than monomers, the MPCD
streaming and collision steps happened infrequently compared to the MD step.

This benchmark demonstrates the strengths of implementing MPCD within an existing
MD package designed to run exclusively on the GPU. We were able to easily perform
simulations with up to 51,200 monomers embedded in 1,250,000 solvent particles on
a single GPU and obtained excellent performance. Moreover, the MD algorithms could
be used as a ``black box'' without special considerations for hiding latency
of migrating MD particle data to the GPU.

\section{Conclusions}
We developed an implementation of the MPCD algorithm for simulating mesoscale
hydrodynamics that is optimized to exploit the massively parallel computational
capabilities of GPUs. Spatial domain decomposition onto multiple GPUs is supported
with MPI, enabling simulations of mesoscale hydrodynamics at length and
time scales that would be otherwise challenging or impossible to access. We showed
that our MPCD implementation efficiently scales to hundreds of GPUs. Here, the
performance was primarily bottlenecked by host-device data transfers, which we speculate
may be mitigated in computing environments having multiple GPUs within a node
and using a CUDA-aware MPI library in conjunction with GPUDirect RDMA. We
found that adopting a mixed-precision computing model for the MPCD particle data
improved performance on a single GPU with acceptable numerical accuracy. We also
showed for a benchmark polymer solution how MPCD can be used to incorporate hydrodynamics
into research-relevant simulations with only modest additional computational cost.
Our MPCD implementation is available open-source as part of the HOOMD-blue project (version 2.3.0)
and should prove useful for conducting simulations of soft matter and complex
fluids when hydrodynamic interactions are important.

In this work, we have focused on implementing the most fundamental components of the
MPCD algorithm. However, there are many extensions of MPCD that are of interest
for areas of active research, including coupling the solvent to solid boundaries such as walls and imposing
external fields to drive solvent flow. We have designed our software to be modular
in order to readily support these extensions. For example, fluid-solid coupling
and flow could be added by modification of the streaming step. We hope to
continue to expand our software with support from the community to incorporate
these and other features in the future.

\textit{Note.}---During the publication process, we determined an improved method for
packing particle data for migration compared to the version that we benchmarked. In the improved version,
we partition the indexes of the particles to be migrated from the particles that are to be retained.
Holes in the particle data arrays left by migrated particles are backfilled
from the list of retained particles to maintain compact arrays, reducing the
number of read and write operations compared to our previous implementation based on stream compaction.
For the SRD benchmark fluid with $L = 200\,a$ on 8 NVIDIA Tesla P100 GPUs, this optimization
reduced the time to pack the MPI buffers by a factor of 12.5x, the overall particle
migration time by a factor of 2.8x, and the total simulation time by a factor of 1.3x. The performance
timings reported in Figures~\ref{fig:weak}-\ref{fig:prof} should also be similarly improved by this
optimization.

\section*{Acknowledgements}
This research is part of the Blue Waters sustained-petascale computing project, which is
supported by the National Science Foundation (awards OCI-0725070 and ACI-1238993)
and the state of Illinois. Blue Waters is a joint effort of the University of
Illinois at Urbana-Champaign and its National Center for Supercomputing Applications.
Additional financial support for this work was provided by the Princeton Center for
Complex Materials, a U.S. National Science Foundation Materials Research Science
and Engineering Center (award DMR-1420541), and the German Research Foundation
under project number NI 1487/2-1.
We gratefully acknowledge use of computational resources supported by the
Princeton Institute for Computational Science and Engineering and the
Office of Information Technology's High Performance Computing Center and
Visualization Laboratory at Princeton University.

\section*{References}
\bibliographystyle{elsarticle-num}
\bibliography{mpcd_gpu}

\begin{thebibliography}{10}
\expandafter\ifx\csname url\endcsname\relax
  \def\url#1{\texttt{#1}}\fi
\expandafter\ifx\csname urlprefix\endcsname\relax\def\urlprefix{URL }\fi
\expandafter\ifx\csname href\endcsname\relax
  \def\href#1#2{#2} \def\path#1{#1}\fi

\bibitem{Praprotnik:2008er}
M.~Praprotnik, L.~D. Site, K.~Kremer, Annu. Rev. Phys. Chem. 59~(1) (2008)
  545--571.

\bibitem{Nagel:2017jm}
S.~R. Nagel, Rev. Mod. Phys. 89~(2) (2017) 025002.

\bibitem{Brady:1993vs}
J.~F. Brady, J. Chem. Phys. 99~(1) (1993) 567--581.

\bibitem{Phung:1996wh}
T.~N. Phung, J.~F. Brady, G.~Bossis, J. Fluid Mech. 313 (1996) 181--207.

\bibitem{Winkler:2013kx}
R.~G. Winkler, S.~P. Singh, C.-C. Huang, D.~A. Fedosov, K.~Mussawisade,
  A.~Chatterji, M.~Ripoll, G.~Gompper, Eur. Phys. J. Special Topics 222~(11)
  (2013) 2773--2786.

\bibitem{Whitesides:2002uq}
G.~M. Whitesides, B.~Grzybowski, Science 295~(5564) (2002) 2418--2421.

\bibitem{Grzybowski:2009ke}
B.~A. Grzybowski, C.~E. Wilmer, J.~Kim, K.~P. Browne, K.~J.~M. Bishop, Soft
  Matter 5~(6) (2009) 1110--1128.

\bibitem{Padding:2006uz}
J.~T. Padding, A.~A. Louis, Phys. Rev. E 74~(3) (2006) 031402.

\bibitem{Allen:1991}
M.~P. Allen, D.~J. Tildesley, {Computer Simulation of Liquids}, {Oxford
  University Press}, {New York}, 1991.

\bibitem{Frenkel:2002chF}
D.~Frenkel, B.~Smit, {Understanding Molecular Simulation}, 2nd Edition,
  Academic Press, San Diego, 2002.

\bibitem{Boger:1977vy}
D.~V. Boger, J. Non-Newtonian Fluid Mech. 3~(1) (1977) 87--91.

\bibitem{Giesekus:1982vs}
H.~Giesekus, J. Non-Newtonian Fluid Mech. 11~(1--2) (1982) 69--109.

\bibitem{Chen:1998vva}
S.~Chen, G.~D. Doolen, Annu. Rev. Fluid Mech. 30~(1) (1998) 329--364.

\bibitem{Ladd:2001gv}
A.~J.~C. Ladd, R.~Verberg, J. Stat. Phys. 104~(5/6) (2001) 1191--1251.

\bibitem{Duenweg:2009ie}
B.~D{\"u}nweg, A.~J.~C. Ladd, {Lattice Boltzmann Simulations of Soft Matter
  Systems}, in: C.~Holm, K.~Kremer (Eds.), {Advanced Computer Simulation
  Approaches for Soft Matter Sciences III}, Vol. 221 of {Advances in Polymer
  Science}, Springer, Berlin, 2009, pp. 89--166.

\bibitem{BRADY:1988up}
J.~F. Brady, G.~Bossis, Annu. Rev. Fluid Mech. 20~(1) (1988) 111--157.

\bibitem{Kumar:2010hy}
A.~Kumar, J.~J.~L. Higdon, Phys. Rev. E 82~(5) (2010) 051401.

\bibitem{Hoogerbrugge:1992hl}
P.~J. Hoogerbrugge, J.~M. V.~A. Koelman, Europhys. Lett. 19~(3) (1992)
  155--160.

\bibitem{Groot:1997du}
R.~D. Groot, P.~B. Warren, J. Chem. Phys. 107~(11) (1997) 4423--4435.

\bibitem{Espanol:2017wy}
P.~Espa{\~n}ol, P.~B. Warren, J. Chem. Phys. 146~(15) (2017) 150901.

\bibitem{Bird:1963ef}
G.~A. Bird, Phys. Fluids 6~(10) (1963) 1518--1519.

\bibitem{Bird:1970bf}
G.~A. Bird, Phys. Fluids 13~(11) (1970) 2676--2681.

\bibitem{Malevanets:1999wa}
A.~Malevanets, R.~Kapral, J. Chem. Phys. 110~(17) (1999) 8605--8613.

\bibitem{Gompper:2009is}
G.~Gompper, T.~Ihle, D.~M. Kroll, R.~G. Winkler, {Multi-Particle Collision
  Dynamics: A Particle-Based Mesoscale Simulation Approach to the Hydrodynamics
  of Complex Fluids}, in: C.~Holm, K.~Kremer (Eds.), {Advanced Computer
  Simulation Approaches for Soft Matter Sciences III}, Vol. 221 of {Advances in
  Polymer Science}, Springer, Berlin, 2009, pp. 1--87.

\bibitem{Bolintineanu:2014dj}
D.~S. Bolintineanu, G.~S. Grest, J.~B. Lechman, F.~Pierce, S.~J. Plimpton,
  P.~R. Schunk, Comput. Part. Mech. 1~(3) (2014) 321--356.

\bibitem{Kapral:2008te}
R.~Kapral, {Multiparticle Collision Dynamics: Simulation of Complex Systems on
  Mesoscales}, in: S.~A. Rice (Ed.), Advances in Chemical Physics, Vol. 140,
  John Wiley {\&} Sons, Inc., Hoboken, New Jersey, 2008, pp. 89--146.

\bibitem{Allahyarov:2002hq}
E.~Allahyarov, G.~Gompper, Phys. Rev. E 66~(3) (2002) 036702.

\bibitem{Ihle:2003cn}
T.~Ihle, D.~M. Kroll, Phys. Rev. E 67~(6) (2003) 066706.

\bibitem{Ripoll:2005ev}
M.~Ripoll, K.~Mussawisade, R.~G. Winkler, G.~Gompper, Phys. Rev. E 72~(1)
  (2005) 016701.

\bibitem{Malevanets:2000tm}
A.~Malevanets, J.~M. Yeomans, Europhys. Lett. 52~(2) (2000) 231--237.

\bibitem{Mussawisade:2005tj}
K.~Mussawisade, M.~Ripoll, R.~G. Winkler, G.~Gompper, J. Chem. Phys. 123~(14)
  (2005) 144905.

\bibitem{Hecht:2005kz}
M.~Hecht, J.~Harting, T.~Ihle, H.~J. Herrmann, Phys. Rev. E 72~(1) (2005)
  011408.

\bibitem{Padding:2005ez}
J.~T. Padding, A.~Wysocki, H.~L{\"o}wen, A.~A. Louis, J. Phys.: Condens. Matter
  17~(45) (2005) S3393--S3399.

\bibitem{Poblete:2014gc}
S.~Poblete, A.~Wysocki, G.~Gompper, R.~G. Winkler, Phys. Rev. E 90~(3) (2014)
  033314.

\bibitem{Lamura:2001un}
A.~Lamura, G.~Gompper, T.~Ihle, D.~M. Kroll, Europhys. Lett. 56~(3) (2001)
  319--325.

\bibitem{Whitmer:2010kp}
J.~K. Whitmer, E.~Luijten, J. Phys.: Condens. Matter 22~(10) (2010) 104106.

\bibitem{Bolintineanu:2012ik}
D.~S. Bolintineanu, J.~B. Lechman, S.~J. Plimpton, G.~S. Grest, Phys. Rev. E
  86~(6) (2012) 066703.

\bibitem{Kanehl:2017fz}
P.~Kanehl, H.~Stark, Phys. Rev. Lett. 119~(1) (2017) 018002.

\bibitem{McWhirter:2009ht}
J.~L. McWhirter, H.~Noguchi, G.~Gompper, Proc. Natl. Acad. Sci. U.S.A. 106~(15)
  (2009) 6039--6043.

\bibitem{Nikoubashman:2014en}
A.~Nikoubashman, N.~A. Mahynski, A.~H. Pirayandeh, A.~Z. Panagiotopoulos, J.
  Chem. Phys. 140~(9) (2014) 094903.

\bibitem{Nikoubashman:2014vh}
A.~Nikoubashman, N.~A. Mahynski, M.~P. Howard, A.~Z. Panagiotopoulos, J. Chem.
  Phys. 141~(14) (2014) 149906.

\bibitem{Howard:2015bl}
M.~P. Howard, A.~Z. Panagiotopoulos, A.~Nikoubashman, J. Chem. Phys. 142~(22)
  (2015) 224908.

\bibitem{Bianchi:2015dx}
E.~Bianchi, A.~Z. Panagiotopoulos, A.~Nikoubashman, Soft Matter 11~(19) (2015)
  3767--3771.

\bibitem{Bianchi:2015corr}
A.~Nikoubashman, E.~Bianchi, A.~Z. Panagiotopoulos, Soft Matter 11~(19) (2015)
  3946.

\bibitem{Nikoubashman:2017bl}
A.~Nikoubashman, Soft Matter 13~(1) (2017) 222--229.

\bibitem{deBuyl:2017hl}
P.~de~Buyl, M.-J. Huang, L.~Deprez, J. Open Res. Softw. 5~(1) (2017) 3.

\bibitem{Petersen:2010cg}
M.~K. Petersen, J.~B. Lechman, S.~J. Plimpton, G.~S. Grest, P.~J. in~'t Veld,
  P.~R. Schunk, J. Chem. Phys. 132~(17) (2010) 174106.

\bibitem{Plimpton:1995fc}
S.~Plimpton, J. Comput. Phys. 117~(1) (1995) 1--19.

\bibitem{Westphal:2014db}
E.~Westphal, S.~P. Singh, C.-C. Huang, G.~Gompper, R.~G. Winkler, Comput. Phys.
  Commun. 185~(2) (2014) 495--503.

\bibitem{Anderson:2008vg}
J.~A. Anderson, C.~D. Lorenz, A.~Travesset, J. Comput. Phys. 227~(10) (2008)
  5342--5359.

\bibitem{Glaser:2015cu}
J.~Glaser, T.~D. Nguyen, J.~A. Anderson, P.~Lui, F.~Spiga, J.~A. Millan, D.~C.
  Morse, S.~C. Glotzer, Comput. Phys. Commun. 192 (2015) 97--107.

\bibitem{Huang:2012ev}
C.-C. Huang, G.~Gompper, R.~G. Winkler, Phys. Rev. E 86~(5) (2012) 056711.

\bibitem{Ihle:2001ty}
T.~Ihle, D.~M. Kroll, Phys. Rev. E 63~(2) (2001) 020201(R).

\bibitem{Ihle:2003bq}
T.~Ihle, D.~M. Kroll, Phys. Rev. E 67~(6) (2003) 066705.

\bibitem{Gotze:2007ec}
I.~O. G{\"o}tze, H.~Noguchi, G.~Gompper, Phys. Rev. E 76~(4) (2007) 046705.

\bibitem{Noguchi:2007fi}
H.~Noguchi, N.~Kikuchi, G.~Gompper, Europhys. Lett. 78~(1) (2007) 10005.

\bibitem{Theers:2015fo}
M.~Theers, R.~G. Winkler, Phys. Rev. E 91~(3) (2015) 033309.

\bibitem{Huang:2015fh}
C.-C. Huang, A.~Varghese, G.~Gompper, R.~G. Winkler, Phys. Rev. E 91~(1) (2015)
  013310.

\bibitem{Huang:2010wt}
C.~C. Huang, A.~Chatterji, G.~Sutmann, G.~Gompper, R.~G. Winkler, J. Comput.
  Phys. 229~(1) (2010) 168--177.

\bibitem{Ihle:2006co}
T.~Ihle, E.~T{\"u}zel, D.~M. Kroll, Europhys. Lett. 73~(5) (2006) 664--670.

\bibitem{Tuzel:2007fr}
E.~T{\"u}zel, G.~Pan, T.~Ihle, D.~M. Kroll, Europhys. Lett. 80~(4) (2007)
  40010.

\bibitem{Tao:2008fk}
Y.-G. Tao, I.~O. G{\"o}tze, G.~Gompper, J. Chem. Phys. 128~(14) (2008) 144902.

\bibitem{Noguchi:2005jz}
H.~Noguchi, G.~Gompper, Phys. Rev. E 72~(1) (2005) 011901.

\bibitem{Wang:2014id}
H.~Wang, S.~Potluri, D.~Bureddy, C.~Rosales, D.~K. Panda, IEEE Trans. Parallel
  Distrib. Sys. 25~(10) (2014) 2595--2605.

\bibitem{Shainer:2011jf}
G.~Shainer, A.~Ayoub, P.~Lui, T.~Liu, M.~Kagan, C.~R. Trott, G.~Scantlen, P.~S.
  Crozier, Comput. Sci. Res. Dev. 26~(3-4) (2011) 267--273.

\bibitem{Potluri:2013by}
S.~Potluri, K.~Hamidouche, A.~Venkatesh, D.~Bureddy, D.~K. Panda, {Efficient
  Inter-node MPI Communication Using GPUDirect RDMA for InfiniBand Clusters
  with NVIDIA GPUs}, in: 2013 42nd International Conference on Parallel
  Processing (ICPP), IEEE, 2013, pp. 80--89.

\bibitem{Colberg:2011de}
P.~H. Colberg, F.~H{\"o}fling, Comput. Phys. Commun. 182~(5) (2011) 1120--1129.

\bibitem{Rapaport:2011el}
D.~C. Rapaport, Comput. Phys. Commun. 182~(4) (2011) 926--934.

\bibitem{Merrill:2011vy}
D.~Merrill, A.~Grimshaw, Parallel Process. Lett. 21~(2) (2011) 245.

\bibitem{cubweb}
http://nvlabs.github.io/cub.

\bibitem{Sutmann:2010jr}
G.~Sutmann, L.~Westphal, M.~Bolten, AIP Conf. Proc. 1281 (2010) 1768--1772.

\bibitem{LeGrand:2013wc}
S.~Le~Grand, A.~W. G{\"o}tz, R.~C. Walker, Comput. Phys. Commun. 184~(2) (2013)
  374--380.

\bibitem{Phillips:2011td}
C.~L. Phillips, J.~A. Anderson, S.~C. Glotzer, J. Comput. Phys. 230~(19) (2011)
  7191--7201.

\bibitem{Salmon:2011eg}
J.~K. Salmon, M.~A. Moraes, R.~O. Dror, D.~E. Shaw, {Parallel Random Numbers:
  As Easy as 1, 2, 3}, in: Proceedings of 2011 International Conference for
  High Performance Computing, Networking, Storage and Analysis (SC '11), ACM,
  2011.

\bibitem{Marsaglia:2000kp}
G.~Marsaglia, W.~W. Tsang, ACM Trans. Math. Softw. 26~(3) (2000) 363--372.

\bibitem{BOX:1958wm}
G.~E.~P. Box, M.~E. Muller, Ann. Math. Stat. 29~(2) (1958) 610--611.

\bibitem{Riesinger:2014wi}
C.~Riesinger, T.~Neckel, F.~Rupp, A.~P. Hinojosa, H.-J. Bungartz, Procedia
  Comput. Sci. 29 (2014) 172--183.

\bibitem{Huang:2010ed}
C.-C. Huang, R.~G. Winkler, G.~Sutmann, G.~Gompper, Macromolecules 43~(23)
  (2010) 10107--10116.

\bibitem{Ripoll:2004jn}
M.~Ripoll, K.~Mussawisade, R.~G. Winkler, G.~Gompper, Europhys. Lett. 68~(1)
  (2004) 106--112.

\bibitem{ieee754}
{IEEE Standard for Floating-Point Arithmetic}, {IEEE Std 754-2008} (2008)
  1--70.

\bibitem{Howard:2017}
M.~P. Howard, A.~Nikoubashman, A.~Z. Panagiotopoulos, {Stratification in Drying
  Polymer-Polymer and Colloid-Polymer Mixtures}, Langmuir (2017)
  http://dx.doi.org/10.1021/acs.langmuir.7b02074.

\bibitem{Zimm:1956ue}
B.~H. Zimm, J. Chem. Phys. 24~(2) (1956) 269--278.

\bibitem{Doi}
M.~Doi, S.~F. Edwards, {The Theory of Polymer Dynamics}, Oxford University
  Press, New York, 1986.

\bibitem{Dunweg:1998va}
B.~D{\"u}nweg, G.~S. Grest, K.~Kremer, {Molecular dynamics simulations of
  polymer systems}, in: S.~G. Whittington (Ed.), {Numerical Methods for
  Polymeric Systems}, Vol. 102 of {The IMA Volumes in Mathematics and its
  Applications}, Springer, New York, 1998, pp. 159--195.

\end{thebibliography}

\end{document}